\documentclass[]{interact}

\usepackage[natbibapa,nodoi]{apacite}
\usepackage{epstopdf}
\usepackage[caption=false]{subfig}

\theoremstyle{plain}

\theoremstyle{definition}

\theoremstyle{remark}

\begin{document}

\title{A study on applications of various Energy Generation in pure Electric Vehicles: progress towards sustainability}

\author{
\name{Dibakar Das, Biplab Satpati\textsuperscript{c},\thanks{\textsuperscript{c}CONTACT Biplab Satpati Email: biplab.satpati@gmail.com}
Md Arif \&
Gourab Das\textit{(Student Member, IEEE)}} 
\affil{Department of Electrical Engineering\\ University Institute of Technology, The University of Burdwan}}

\maketitle

\begin{abstract}
The present work is an attempt to understand and review existing methods of energy generation in electric vehicles in the modern day context. Previous works in the field have proposed various mechanisms of energy generation that are very well adaptable to commercial scale uses and can be used as alternative power sourcing for electric vehicles having nil or very low environmental impact. The paper discusses strategies such as photovoltaic cell systems, regenerative braking, fuel cell, thermoelectric generators and micro wind-turbines with adequate propositions to select them on the basis of their suitability. The document also includes important formulas that can be used for individual modeling and designing. The paper emphasises on introducing the mechanisms that can be introduced as assistive mechanisms or secondary sources so that the range and other parameters are not compromised. 
\end{abstract}


\section{Introduction}
Electric Vehicles are currently designated as the carriers of modern human civilisation but their capabilities are constantly questioned due to constraints such range anxiety, lower sustainability and high maintenance. The most crucial of all the problems that lies in the concern of electrical and energy engineers is the problem of range anxiety in developing countries that lack specific charging infrastructures that are required for mass migration towards automotive electrification. The advent of electric vehicles was inevitable due to the sheer amount of environmental hazard that IC engine vehicles poses, and although phenomenal advances have been made by chemical and mechanical engineers which have reduced emission and increased the efficiency but the current rate of global consumer growth has nullified the efforts. Electric Vehicles presently mostly use Battery EV or BEVs which are powered by charging stations connected to the grid. Any EV while traversing  through a terrain faces a lot of challenges, but the terrain is not always the villain, for example while getting down from a hill the vehicle gathers momentum and faces air resistance. Both of these factors can be used to gain energy if the power-train is suitably designed the generated power can be stored in the battery thus increasing the overall economy. Similarly additional fuel cells can be used to increase the range of the vehicle. In our present scope we have inspected on photovoltaic cell systems, fuel cell systems, regenerative braking systems, thermoelectric generators and wind turbine systems.
\par In the present work we have taken an brief inspection on the various methods that exist in the space of EV and how their application can be fruitful in increasing the overall usability and efficiency of any BEV or pure EV. The silent introduction of the individual methods like solar PV systems in EV can be gradually modified with the advances in researches in the field to exercise coexistent development in a EV that has more and more dependency on the PV generation than the grid, thus providing autonomy. With that in mind, it is also to understand that at present, the technologies are either nascent or have intrinsic limitations which limit their capabilities to be used as a primary energy source. Due to the above stated fact we have emphasised more into the possibilities as secondary generation source rather than being the primary energy source of the EV. 

\section{Photovoltaic Cell Systems}
 Solar or photovoltaic cells have been used to generate electricity from sunlight for a long time. Although individual cells only generate 0.5V to 1V, constructing them into an array may generate a few watts and furthermore constructing it into modules can produce a larger amount of power \citep{kalantar2010dynamic} and a generic power system design has been proposed in \citep{sarkar2014generalized}. In electric vehicles, since the storage is DC the solar PV modules output can be directly stored in the battery by only specific DC-DC converter controlled by a Charge Controller. The Charge Controller ensures uniform charging of the batteries in the battery bank. In specific cases an Inverter may be required to convert the power into AC to be directly fed to the AC motor in the vehicles . The only constraints to solar PV based generation systems are large upfront cost and dependence on weather conditions \citep{deshmukh2008modeling}. PV systems are novel due to the fact that they do not produce any greenhouse gases or pollutants during their operation \citep{wall2013advantages}, but some pollutants are produced during manufacturing \citep{burkhardt2012life}, which however is much lesser than conventional fossil fuels. Solar PV are also advantageous because they have lesser maintenance cost and have no noise pollution \citep{piano2017toward, wall2013advantages}. Electric bicycles are effective in developing countries due to their carrying capacities \citep{patoding2021electric}.
\par The equivalent circuit that can represent the PV model can be given as \citep{villalva2009modeling, villalva2009comprehensive, salmi2012matlab}:
\begin{equation}
I= I_{pv}-I_0 
 \left[\exp  \frac{q(V+R_s I)} {N_sKTa}  \right] - \frac{V+R_sI}{R_p}
\end{equation}
where $N_s$ refers to the number of cells in the module, K is the constant referred to as Boltzman constant, T refers to the model's absolute temperature, $a$ is the ideality factor of the diode; $R_p$ and $R_s$ are the equivalent series and parallel resistance of the module; V denotes the voltage across the diode; $q$ refers to electron charge and; $I_0$ and $I_{pv}$ represent the saturation current and PV saturation current of the module respectively.
\par The relation between PV current $I_{pv}$ and the incident solar radiation and the temperature of the module can be written as:
\begin{equation}
I_{pv} = [I_{pv,n} + K_1 (T-T_n)] \frac{G}{G_n}
\end{equation}
 where $K_1$ indicates the coefficient of short circuit current temperature (A/\textdegree  C); $T_N$ and $T$ denote the the nominal and actual temperature in Kelvin; $G_N$ and $G$ denote the the actual and nominal output in (W/m$^2$); at the nominal conditions of solar radiation at 1000 W/m$^2$ and 25\textdegree ,  the current generated is represented as $I_{pv,n}$.
\par The temperature dependent saturation current can be described by:
\begin{equation}
I_0 = \frac{I_{sc,n}+K_1(T-T_n)}{\exp \frac{q[V_{oc,n}+K_V(T-T_n)]}{aN_sKT}-1}
\end{equation}
where $K_V$ and $K_I$ are the voltage and current coefficients respectively and the open circuit voltage (in V) and short circuit current (in A) are represented by $V_{oc,n}$ and $I_{sc,c}$, respectively.
\par Solar cells have been used in Solar cars to power the whole system with modern energy management strategies such as Maximum Power Point Control using Control Area Network can achieve very high efficiency and viability \citep{babalola2021solar, taha2008review, mangu2010design}. PV has to be tested for efficiency in a automobile before reacheaing the roads and various methods have been designed to achieve the same using computational methods are different from rooftop solar PV modules \citep{schuss2021advanced},  in \citep{jeddi2017comparative} it has been shown that a two diode model is the best fit for simulational studies. Broadly solar PV can be classified into first, second and third generation. The first generation uses monocrystalline or multicrystalline/polycrystalline silicon based heavy fixed and non flexible solar PVs but they are not of much importance now due to their drawback for high price and low efficiency.
\begin{figure}[ht]
    \centering
    \includegraphics[width=0.40\textwidth]{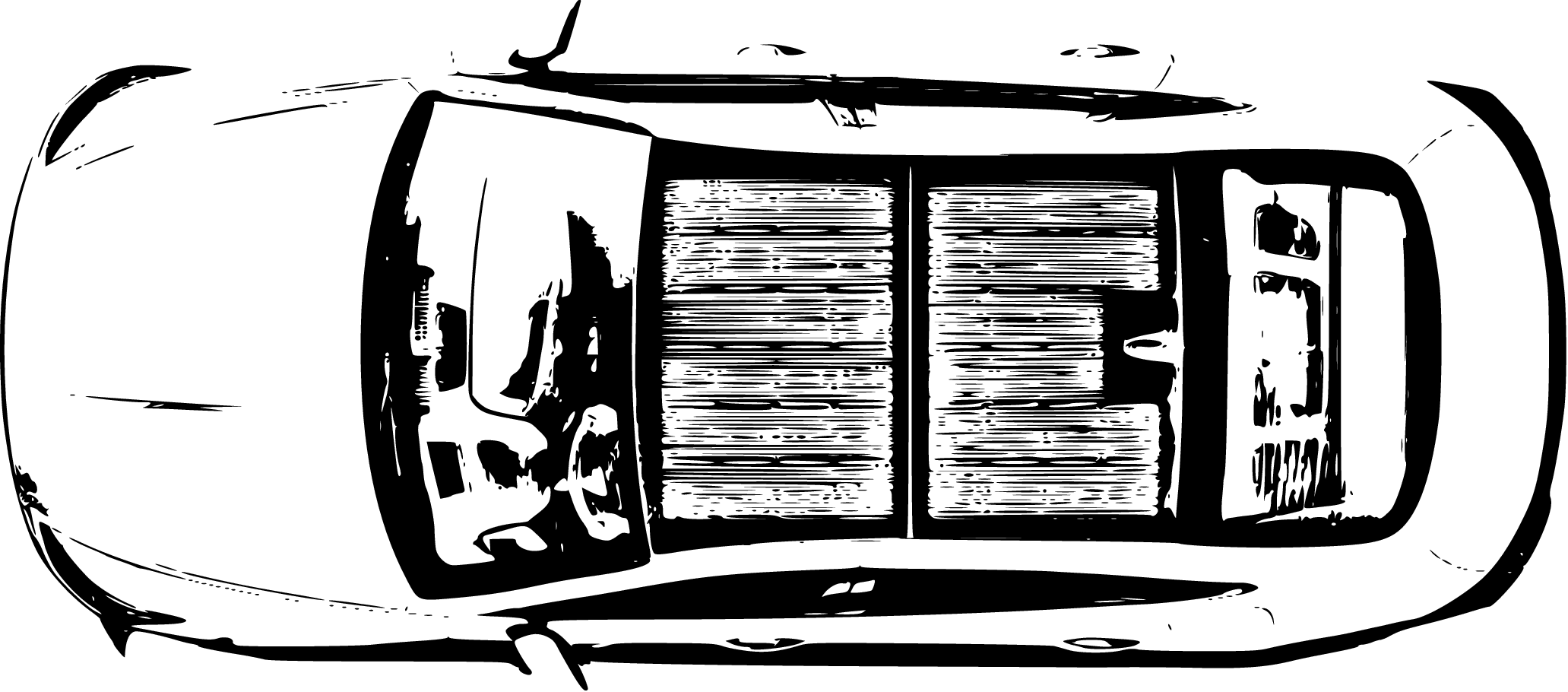}
    \caption{Sketch diagram of Hyundai Car with solar roof charging \protect\citep{hyundai}.}
    \label{3.1}
\end{figure}

The most fundamental problem of solar PV systems is the sheer low rate of  efficiency and mould-ability on vehicle bodies. Second generation PV uses amorphous silicon, silicon polymers or non-silicon materials that have lower weights, higher efficiency and are economically viable. In this category 28.0\% for perovskite/Si dual-junction tandem solar cells, 32.8\% efficiency for III-V/Si dual-junction and 35.9\% efficiency for III-V/Si triple-junction is evident and can achieve over 40\% efficiency in theory in certain conditions, other than this properties like flexible, coloured PV with static concentrator design makes them suitable for automobile applications \citep{yamaguchi2020role}. With advancing technology solar PV has been designed to be of high efficiency and light weight to be ideally mounted on automobiles. In \citep{kraiem2021increasing} a vehicle had design had been proposed with upper and sides of the vehicle had been covered with mounted solar and Particle swarm optimization has been used to provide higher output in terms of maximum power point tracking but fundamentally these designs are neither sustainable nor marketable and the most important requirements of innovation lie in the domain of electronics design more than power controller design domains. 
\par Vehicle integrated photovoltaics (VIVP) is a family of photovoltaic that is used on vehicles and have lighter weights than conventional PV modules are highly efficient specially due to their concentrator photovoltaic module \citep{kim2021future, ekinspotential,brito2021urban} and their scope has been further explained with analysis reducing charging time and improved cooling with sustainable cost to be estimated for near future \citep{heinrich2020potential}. The potential disadvantages are requirements for highly sophisticated energy management and higher losses \citep{heinrich2020potential,mahmoudi2014overview}. Modelling in VIPV has to be done very carefully as their shape and positioning is way different than traditional PV systems \citep{araki2020measurement} and as a standard 1m radius of curvature with 96\% coverage has been suggested \citep{ota2022facilitating}. These PV requires major changes in structural, thermal and physical properties to suit the requirements \citep{nukunudompanich2022aspects}.
\par The most advanced PV cells are made by cutting edge technologies such as printed, dyed or plastic based solar cells. Recently  inkjet printer based solar PV manufacturing architecture has been made, it has greater freedom of design and is cheaper than previous variants \citep{karunakaran2019recent, brunetti2019printed,wu2019slot,cheng2016print}; if solar PV printing could be executed at larger  scales the possibility of this technology is huge.
\section{Fuel Cell Systems}
Fuel cells (FC) serve as a source of clean energy with almost no carbon footprint and are considered important for mitigating lack of infrastructure leading to lack of charging stations at remote places \citep{muthukumar2021development, wilberforce2017developments,srinivasan1981fuel}. In multiple models, FC have been used in HEVs as well as PEVs. But mono-applicable energy generation by FC is generally not supported as they need special refuelling stations which are still not present in abundance. Various different types of Fuel Cell have been used and thus there are serious contenders in the innovational scope \citep{mekhilef2012comparative}. 
The ideal voltage generation for a FC is derived from the Nernst equation  described as follows:
\begin{equation}
E_{cell}=E_0 + \frac{RT}{2F} \ln \frac{P_{H_2}{\sqrt{P_{O_2}}}}{P_{H_{2}O}}
\end{equation}
where $E_{cell}$ represents cell output, $T$ is the absolute temperature of the cell, $R$ is universal molar gas constant in which the cell is working, $F$ is Faraday's constant, the partial pressure of gas water, oxygen and hydrogen can be represented as $P_{H2O}$, $P_{O2}$ and $P_{H2}$ respectively. But the individual cell voltage is less than ideal voltage due to resistive losses, internal current losses, concentration losses and activation losses. Thus practical equation that can be used to find the output voltage of the SC stack can be described as:
\begin{equation}
V_{FC}= N \bigg( E_0 + \frac{RT}{2F} \ln \frac{P_{H_2}{\sqrt{({{P_{O_2}}/{P_{std}})}}}}{P_{H_{2}O}} -V_L \bigg)
\end{equation}
where $V_L$ is the voltage losses in individual cells that was described in the above segment, $P_{std}$ is the standard pressure, and $N$ is the number of cells that are present in the stack. 
\par Direct Hydrogen FC are better than alcohol based Hydrogen generation as they are cleaner and can also be mass produced \citep{thompson2018direct,tancc2019overview} and have great scope for improvement in the coming decade, the reduction in their cost and and viability is projected to improve their importance in the coming times. A general simulation model was provided in \citep{moore2005dynamic} which is still relevant in designing Hydrogen FC systems. 

\par PEMFC is an fuel cell that has been extensively used for applications in EVs \citep{henao2012proton, fathabadi2019combining} and was suggested in \citep{swan1991proton, swan1994proton} and various membranes in various applications was analsyed in \citep{wakizoe1995analysis}, the models of use and simulation was presented in detail and used numerously due to its accuracy in developing the basic modelling \citep{boettner2002proton}. The performance analysis of PEMFC shows its success in urban drive cycle which still now encompasses the highest vehicle density of the world and will continue to do so in coming times \citep{farhani2021experimental, ogungbemi2021selection}. The application of PEMFC is also getting enriched by continuous research in control schemes and specific driving conditions that are essential in increasing the overall efficiency and  FUZZY logic- classical PI configuration and  state machine strategy outperform others in various metrics \citep{soumeur2020comparative}. EVs with SOFCs (solid oxide fuel cells) have been designed as a range extender system in \citep{dimitrova2017environomic} increasing economy by over 600 km and various modelling schemes to design and analyse the parameters of SOFC were presented in \citep{yang2020state}. SOFC enjoys the benefit of using solid fuel and thus are easy to carry and also have less electrolyte loss Maintenance but they also have disadvantages such as supply characteristics which are not uniform and linear and also has higher material maintenance; the supply characteristics which was corrected in \citep{rafikiran2022design}. PEMFCs (Proton Exchange Membrane Fuel Cell or Polymer Electrolyte Membrane Fuel Cell) is a better alternative to SOFCs due to low operating temperatures \citep{sorlei2021fuel, baba2021fuel}.  

\par Other types of FC are: direct alcohol fuel cell (DAFC), phosphoric acid fuel cell (PAFC), alkaline electrolyte fuel cell(AFC) and molten carbonate fuel cell (MCFC), but they have lower applicability or efficiency than both SOFCs and PEMCFs \citep{amba2022direct, park2021economic,li2020relating}. DAFC uses alcohol for production of hydrogen in the FCs and mechanisms like electrocatalysis is used and their application in EVs were also suggested \citep{lamy2001electrocatalytic, lamy2002recent, sen2005investigation, vaidya2006insight, vigier2006electrocatalysis}. Alkaline electrolyte FC have also been proposed to be incorporated in EVs and they have much lower operating cost than other contemporary applications like PEMFCs but require higher maintenance cost \citep{mclean2002assessment} and  their initial innovation was recorded in \citep{kordesch1999intermittent}. The lack of popular innovation in the field highlights the shortcomings of AFC. PAFC has been assimilated in pick-up EVs to improve range \citep{pathak2006development} and various matching conditions optimal to operation or various concentration and other physical factors in association with load resistance has been observed in \citep{srinivasan1981fuel}. MCFC has been integrated to be used with thermoelectric generators to simultaneously improve the power density and efficiency providing the critical hydrogen required for operation along with heating and cooling generally referred to as trigeneration \citep{chen2022performance, li2013analysis}.
\par FC based electric vehicles face issues of durability and fault diagnosis but with innovations in durability, reliability and storage the use of FC in electric vehicles as a sparing energy source may become important in the coming future \citep{inci2021review} and advanced optimisation based control strategies already has significant outcomes in terms of reduced power fluctuations improving the economy \citep{gharibeh2020energy, lu2020energy} and similar research was performed in optimal control by quadratic programming \citep{lin2021real} improving the SoC characteristics of the EV battery. FCs have also been used in conjugation with or independently with batteries to power two-wheel vehicles \citep{bui2021energy} and a FC-supercapacitor based recreational vehicle was presented in \citep{macias2021fuel} and passive topology shows most efficient system and provides a possibility for a future scope. 
\par FC system incorporated EVs also provide better performance in PV integrated EVs or VIPV systems that are governed  by MPPT systems \citep{he2020novel}. The future of FC PEVs has the possibility of onboard Hydrogen generation technologies that reduce the problems arising of storage and refuelling \citep{shusheng2020research}.

\section{Regenerative Braking Systems}
Energy once lost can hardly be recovered so it is in best of interest to convert the waste energy into reusable forms of energy, in electric vehicles this has been essentially done by a framework of regenerative braking. Initially regenerative braking was used for large sized transports as the conversion efficiency was lower as well as overall energy generated by classical means were very low as compared to the system requirements. Electric vehicles were supported by majorly four types of regenerative braking systems: Spring and Flywheel in mechanical and Supercapacitor and Battery storage based in Electrical.
\subsection{Mechanical Systems}
The cost compatibility has been solved by using spring regenerative braking systems which store the energy as potential energy, the only dominant problem is that they have very low energy efficiency \citep{jiang2013research}. In many electric vehicles the braking energy gets stored as  kinetic energy in the flywheel arrangements.
Spring mechanical systems can be used to convert kinetic energy in a braking wheel and store it in coil springs to be used later for acceleration or for providing starting torque \citep{myszka2015mechanical}. Much research has been done on spring based regenerative braking in HEVs but much lower has taken place for PEVs but in recent studies spring based systems have been designed to convert the mechanical power directly into electrical energy with 53\% efficiency \citep{qi2020electro}.

\par Flywheels like springs are mechanical devices that can be used to store excess energy during braking but unlike springs, flywheels store energy in the form of kinetic energy \citep{itani2016energy}. Flywheels had been used as energy storage elements in electric vehicles \citep{daberkow2013electric} but their use as storage elements during regeneration is unique. Flywheel based RBS can save 25\% of the total vehicle power during its operation \citep{budijono2019development}. 
Flywheel can achieve upto 25\% improved less fuel consumption efficiency, but this requires the flywheel to be very well designed and made of material of very high material strength \citep{bhanewaddesign}. The use of advanced optimization techniques in controlling the Flywheel energy systems using Neural Network models have been used to improve performance in RBS \citep{wang2021optimization}. 

\subsection{Pure Electrical Systems}
Supercapacitors are energy retention devices that can temporarily store energy during braking; this power can be used while subsequent driving \citep{naseri2016efficient} the Supercapacitors also have additional functions like higher acceleration and battery safety. The SCs have been shown to provide almost 68\% peak energy efficiency and thus reliable but SCs are not without limitations one such is use of converters which cannot essentially match the charging and discharging cycles of the SCs \citep{partridge2019role}
\par Supercapacitors were used to store the energy generated during braking by Motor/Generator couples, their use with supercapacitors to store the produced electricity was prominent earlier, but although they have great energy density they are expensive and costly to maintain \citep{li2019comprehensive} and they have lower . Later the supercapacitors were replaced and complex converters were introduced in the powertrains to return power back to the energy storage system (ESS). Supercapacitors (Ultracapacitors) show better performance than FES (flywheels) in domains such as multiple discharge drive cycles, weight and specific energy while they are lower in performance in economic and volumetric constraints, energy and power density \citep{itani2017comparative}.
\par The classical motor/generator couple had a lot of problems that contributed to its low efficiency, generally DC motors had to be used with this kind of braking systems as conversion was uneconomical. With the advent of power-electronics and research to improve its functioning the motor/generator could be used to generate power while breaking. The power would then travel through the similar lines through which the operational motor power came through. This income and outflow of power is generally maintained by flyback-converters, cycloconverters or matrix converters \citep{gayathri2021power}. In advanced new age EVs generally other techniques are not used very frequently and now EVs rely solely on power electronics based systems and stores the energy in onboard energy storage devices \citep{xu2015fully,zhang2018research}. Recent work has also visited sliding braking by switched reluctance motor drive system \citep{zhu2021regenerative} and bidirectional DC-DC converter for powering permanent magnet DC motors with almost efficiency twice as that of conventional schemes \citep{kumar2021modified}. In \citep{kumar2022regenerative} a customised quadratic gain bidirectional control has been developed to achieve the desired qualities of regenerative  braking with a special converter topology with efficiency reaching upto 95\%. Flywheels have been also used with BTB i.e.back to back converters to achieve higher efficiency  \citep{babu2020analysis} but they have undesirable characteristics in terms of weight and volume \citep{erhan2021prototype}. 

\par Analysis shows that regenerative braking has a good effect on battery lifetime and braking performance in automatic braking systems (ABS). Modern regeneration has achieved higher efficiency by using control and optimization techniques without affecting the quality of ride of the passengers; they have been presented briefly in \citep{jamadar2021review}. Regenerative braking has also been used extensively in e-bikes and scooters \citep{chandru2021design, ilahi2020design, schneider2021instant}. Even in hybrid electric vehicles that are designed towards Through-the-road architecture the regeneration could be as high as 40\% by modulating the deceleration speeds without compensating safety \citep{rizzo2021optimal} and optimised modelling of braking force distribution can improve the performance upto 51.9\% without compensating the braking safety \citep{biao2021regenerative}.
\par The most important and advanced aspect of regenerative braking can be seen as autonomous swarm driving based on car following that enables the system to pre-calculate the stoppage time and other parameters and the proposed model delivers 57.61\% more energy efficiency than non-regenerative braking strategies \citep{guo2021safe}. Additionally \citep{li2021energy} shows that autonomous vehicles can also work superlatively due to constraints that are introduced while designing the controllers of the converters that operate during the regenerative braking by removing the non-linearities in motion introduced by drivers.

\section{Thermoelectric Generators}
Thermoelectric generators or thermogenerators are devices that increase the overall efficiency of the electric vehicles by converting heat energy produced in the electric vehicle by converting into alternating forms of energy easily by use of pipes and TEGs \citep{orr2016review} and in general pure electric vehicles do not employ this as they work more efficiently in ICE engine based Hybrid EVs but recently the use of thermoelectric generations have been revisioned for braking energy recovery specially for range extended EVs \citep{lanmatching, hewawasam2020waste} but our concerns is more limited to pure EVs so deeper analysis in this aspect has not been done. There have been efforts made to use TEGs in PEVs as well \citep{tatarinov2013modeling} and their role can be significant in long terms to achieve sustainability \citep{anandasabesan2021work}.
\par Thermoelectric generators can be made by silicon, ceramics and polymers \citep{he2015recent, jaziri2020comprehensive, van2020thermoelectric} and the optimal number of cells required to gain maximum power and minimise the resistance power consumption \citep{champier2017thermoelectric, ge2022experimental}. The importance of optimisation and matching while selection in EVs has been extensively studied and the necessity of optimisation has been highly emphasised \citep{lan2022matching}. Thermogenerators generally are not employed due to their low efficiency as their energy losses may surpass their energy savings \citep{fernandez2021thermal} but intelligent vehicle cooling systems may balance out energy consumption costs by merging with exhaust facilities thus both decreasing energy losses and improving the efficiency of cooling as well as the thermogenerator efficiency \citep{bencs2021potential, pacheco2020compact}. In \citep{omar2020motorcycle} thermoelectric generators have been used to analyse the viability of thermoelectric generators in motorbikes, the generated energy can be used for low power application.
\par Thermoelectric generators can be modified to support auxiliary instruments in electric vehicles that may be generated by braking or heating of power handling devices \citep{coulibaly2021use}. Braking is an important field of application that employs TEGs to generate power, the brake drums gets heated due to the friction with the brake shoes to generate heat, this heat can be reused by TEGs to generate upto 4W power to support auxiliary instruments \citep{coulibaly2021use}. Proposals have been made to employ similar mechanisms in electric buses to use braking heat which can go as high as 259.47 °C for moving trucks \citep{siregar2021drum}.
\par Recently the use of TEG has been proposed in \citep{ramadhani2020optimal} in Fuel cell EVs improving the heat recovery mechanism and reducing 12.85\% consumption of grid power and have also been used with Solar PV to enhance the PV performance by increasing the efficiency by providing partial cooling to the PV cells which can work optimally ast lower temperatures due to reduced thermal runaway losses \citep{moh2017hybrid}. Perhaps the most innovative advances have been made towards integrated energy recovery mechanisms in batteries \citep{pourkiaei2019thermoelectric, lu2020research, tu2022comprehensive} and they have a promising future but their cost efficiency at present is not highly efficient for EV applications. In specially situations that require drivetrains in tropical scenarios TEG based Battery Management Systems work the best in cooling the battery and simultaneously gathering some power \citep{li2019experimental, kim2019review}.

\section{Wind Turbine systems}
Wind turbines are an alternative source of renewable energy and can act as mobile energy generation systems; but their application in electric vehicles have been gathering interest from the first decade of the 21st century \citep{boodman1982wind, deets2006wind}. Initially this began with the idea of simply mounting a wind turbine referred to as vehicle mounted wind turbine (VMWT) on an EV either vertically or horizontally\citep{shivsharan2020review}. Later various techniques were introduced to incorporate better and efficient wind turbines in electric vehicles with varying degrees of design and mechanical innovations most important to which include the use of savonius rotor wind turbine (SRWT) which are placed in hollow enclosures or mountings. Present technologies use designed and dedicated vents in EVs to derive power during motion. Even later turbines were designed which were designed similar to turboshaft and turboprop turbines similar in construction to those used in aircrafts. Comparative analysis show that wind turbines show least effects to drag and aerodynamic performance if they are placed in front bumper of the EV as compared to hood or rooftop \citep{sofian2014evaluation}. The cases of VMWT application and their mathematical analysis in electric vehicles have been shown in \citep{Rubio_Llopis-Albert_2019} and shows the power enhancement in terranous drive cycles.
\par First generation VMWT have a major disadvantage i.e. they pose higher drag losses in the vehicles and nullify the energy generation by the turbines. The vertical axis wind turbine (VAWT) \citep{awal2015design, jiang2016preliminary, weng2018investigation, hussain2021charging} has serious drag effects and has a very incohesive design. The positioning of the rotor behind the roof console disbalances the turbine blades which are outcast by the console during the movement; these small inconsistencies may drastically reduce the performance. The number of blades can be calculated by computational fluid dynamics (CFD) and has been studied in \citep{shivsharanmodal} and can be extended for application in EVs providing that three, five and eight blades can be used for these applications. Horizontal axis VMWT have a better compatibility than the vertical axis VMWT but their applications are also limited due to being their bulky nature or higher drag \citep{bao2012experimental}. Other than this there are turboshaft or turboprop arrangement turbines that have been used \citep{quartey2014generation} but they aren't favorable due to their positioning as they increase drag losses as previously stated \citep{sofian2014evaluation}.
\par A second generation technology uses Savonius Rotor Wind Turbine (SRWT) i.e. Savonius Rotor Vertical Axis Wind Turbine (SR-VAWT) \citep{arbiyani2020design, hussain2021charging} and Savonius Rotor Horizontal Axis Wind Turbine (SR-HAWT) \citep{awal2015design} and its construction has been suitably modified for commercially produced vehicles. The design of the turbine is a factor of the speed of the vehicle and thus the blade area requirement criterion can pose a problem, but the overall construction is sustainable and can be practically used for commercial vehicles. A 3-D printed model has been shown in \citep{monzamodeth2021development} which has been designed for speeds less than 80 km/h but the construction is not highly standardized due to its bulky and unappealing looks and problems in material properties.
\par A significant work has been done in \citep{feudjio2022optimization} and the most optimal configuration and position to be selected for mounting the turbine on the vehicle incorporating SR-HAWT has been recognised as the frontal position with straight generators enumerating it as the most effective position for placement of SRWT on EVs and detailed aerodynamic analysis in suggest that turbines shall be attached with straight generators and not side generators \citep{fotso2019aerodynamic}. In general two, three or six blade SRWT are used \citep{alom2018four}, but by using methods like increasing the number of blades to six and removing the interference of wind on convex parts the power output of the turbine can be increased by up to 60\% ie from 0.3 to 0.5 \citep{emmanuel2011numerical, alom2018four} and in case of two blade  SR-VAWT the blades need not be placed completely opposite to each other rather each blade should overlaps a small part of the opposite blades increasing the effective area under contact \citep{olaoye2016numerical}.
\begin{figure}[ht]
    \centering
    \includegraphics[width=0.4\textwidth]{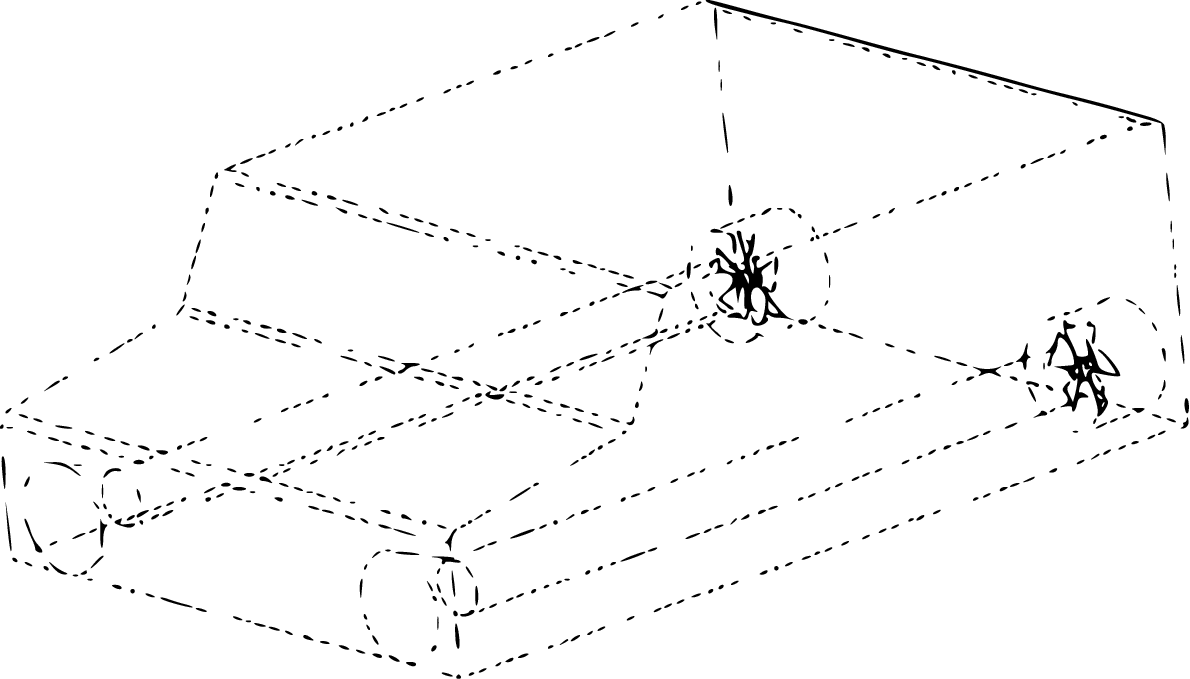}
    \caption{Isometric view as designed in \protect\citep{gupta2021energy}.}
    \label{3.2}
\end{figure}
\par The most advanced type of wind turbine system uses a duct system that houses the turbine within the duct and can be referred to as Duct housed Wind Turbine (DWT) \citep{rosly2017flow, gupta2021energy}, the air moves into the duct while in motion and applies thrust to the built in turbine to generate auxiliary power, this type of energy generation or harvesting can also be called as aerodynamic energy harvesting and design of DWT has to be done according to frontal design of the EV so as to increase its efficiency \citep{solanki2021aerodynamic}, this technique unlike previous methods do not place additional structures outside the EV rather places it so as to facilitate incoming winds into the duct. This type of DWT can contain both SRWT or turboshaft/turboprop type turbine structure. If the issues are resolved micro-wind turbines are an important innovation in solving the energy crisis and its potential has been seen even in commercial sized buses \citep{chellaswamy2017future}. In recent times a Wind Powered EV, Eolo has been designed by Colombian engineers and its academic analysis has been presented in \citep{garcia2021performance} and similar works can be done so as to improve commercial vehicles. Presently e-bikes have also been proposed that are compatible with wind turbine based generation systems \citep{kumar2020experimental, chevli2020development, shivsharan2021experimental}. 
\par VAWT with vent based systems can be used to increase the system's overall \citep{jiang2016preliminary} and increase the EVs energy efficiency but continuous duct based wind turbine is more sustainable and appealing and requires more research attention but DWT can also be collaborated with other renewable sources like solar or FCs to provide power to EVs and their possibility in the future is enormous \citep{mamun2022systematic}. An innovative approach has been observed for electric powered pickup trucks with SR-VAWT and these gradual innovations prve way to reach sustainability \citep{mohd2016proposed, mohd2016aerodynamic} the design and mechanism of the housing or enclosure along with the SR-VAWT compliments the structure and construction of the hood of the truck thus making it relevant to be relooked and reinvestigated upon in coming times.

\section{Conclusion}
The importance of power generation strategies have been reviewed in scale and breadth in the paper and various previous works have been assimilated to enrich the technological advancements in the field. Various systems like solar PV systems can  be easily used for energy generation, even in sophisticated EVs without loss of visual appeal. Regenerative braking has been used for a long time as an energy re-usage mechanism and its gradual transform has made its acceptance universal in all forms of electric vehicles. Fuel cell systems can be used to increase the range of EVs and can be used as replacement for battery in electric vehicles. In addition to these widely used technologies additional power generation schemes like Thermo-elecetric generators and micro wind-turbines can be introduced to use the auxiliary power that would otherwise have been wasted into the environment. These technologies improve the scope for increasing the efficiency of electric vehicles and improving their viability and sustenance over long time and may provide as the starting point for energy-independent automotive.
\par The methods call for use of additional conversion or generation structures in addition to the conventional powertrain components, but regenerative braking can be introduced with very little modification to the powertrain of the EV, thus each method needs to be analysed on a cost-benefit scale and could be addressed in a later work.

\end{document}